\documentclass[aps,pra,twocolumn,superscriptaddress,floatfix,longbibliography]{revtex4-2}

\usepackage[utf8]{inputenc}
\usepackage[T1]{fontenc}
\usepackage{bm}
\usepackage[english]{babel}	
\usepackage{breakcites}
\usepackage{lmodern}
\usepackage[normalem]{ulem}

\usepackage{ae,aecompl}
\usepackage{subfigure}
\usepackage{latexsym,stmaryrd}
\usepackage{graphicx}
\usepackage[labelfont=bf,font={small},labelsep=period]{caption}
\usepackage[hidelinks,linkcolor=blue,colorlinks=true,urlcolor=blue,citecolor=blue]{hyperref}
\usepackage{braket}
\usepackage{amsmath}
\usepackage{upgreek}
\usepackage{cleveref}
\usepackage{comment}
\usepackage{url}
\usepackage{amssymb}

    \newcommand{\ac}{\alpha}
\newcommand{\ncav}{\overline{n}_\mathrm{cav}}
\newcommand{\ain}{\ac_\mathrm{in}}
\newcommand{\aout}{\ac_\mathrm{out}}

\newcommand{\pin}{P_\mathrm{in}}

\begin{document}

\title{Intermodulation of optical frequency combs in a multimode optomechanical system}

\author{Ryan C. Ng}
\thanks{These authors contributed equally to this work}
\affiliation{Catalan Institute of Nanoscience and Nanotechnology (ICN2), Campus UAB, Bellaterra, 08193 Barcelona, Spain}

\author{Paul Nizet}
\thanks{These authors contributed equally to this work}
\affiliation{Catalan Institute of Nanoscience and Nanotechnology (ICN2), Campus UAB, Bellaterra, 08193 Barcelona, Spain}

\author{Daniel Navarro-Urrios}
\affiliation{MIND-IN2UB, Departament d’Electrònica, Facultat de Física, Universitat de Barcelona, Martí i Franquès 1, Barcelona 08028, Spain}

\author{Guillermo Arregui}
\affiliation{DTU Electro, Department of Electrical and Photonics Engineering, Technical University of Denmark, {\o}rsteds Plads 343, DK-2800 Kgs. Lyngby, Denmark}

\author{Marcus Albrechtsen}
\affiliation{DTU Electro, Department of Electrical and Photonics Engineering, Technical University of Denmark, {\o}rsteds Plads 343, DK-2800 Kgs. Lyngby, Denmark}

\author{Pedro D. Garc\'{i}a}
\affiliation{Catalan Institute of Nanoscience and Nanotechnology (ICN2), Campus UAB, Bellaterra, 08193 Barcelona, Spain}

\author{S{\o}ren Stobbe}
\affiliation{DTU Electro, Department of Electrical and Photonics Engineering, Technical University of Denmark, {\o}rsteds Plads 343, DK-2800 Kgs. Lyngby, Denmark}
\affiliation{NanoPhoton - Center for Nanophotonics, Technical University of Denmark, {\O}rsteds Plads 345A, DK-2800 Kgs.\ Lyngby, Denmark}

\author{Clivia M. Sotomayor-Torres}
\affiliation{Catalan Institute of Nanoscience and Nanotechnology (ICN2), Campus UAB, Bellaterra, 08193 Barcelona, Spain}
\affiliation{ICREA - Instituci\'o Catalana de Recerca i Estudis Avan\c{c}ats, 08010 Barcelona, Spain}

\author{Guilhem Madiot}
\thanks{Corresponding author: guilhem.madiot@icn2.cat}
\affiliation{Catalan Institute of Nanoscience and Nanotechnology (ICN2), Campus UAB, Bellaterra, 08193 Barcelona, Spain}

\begin{abstract}
Phonons offer the possibility to connect the microwave and optical domains while being efficiently transduced with electronic and optical signals. 
Here, we present a multimodal optomechanical platform, consisting of a mechanical-optical-mechanical resonator configuration. The mechanical modes, with frequencies at 265 MHz and  6.8 GHz, can be simultaneously excited into a phonon lasing regime as supported by a stability analysis of the system. Both the MHz and the GHz modes enter a self-sustained oscillation regime, leading to the intermodulation of two frequency combs in the optical field.
We characterize this platform experimentally, demonstrating previously unexplored dynamical regimes. These results suggest the possibility to control multiple mechanical degrees of freedom via a single optical mode, with implications in GHz phononic devices, signal processing, and optical comb sensing applications.

\end{abstract}

\maketitle

\section{Introduction}

Phonons exist over an extremely broad frequency range from the Hz to THz regime. The control of phonons is difficult, especially at higher frequencies, despite the implications that such control would enable. For example, in the GHz regime, this could enable phonon signal processing, nano-acoustic devices, or hybrid quantum systems such as superconducting qubits \cite{VanLaer2015,Kittlaus2021,Krantz2019}. Recent advances in the realization of high-frequency nano-acoustic devices have further motivated the extension of concepts from electronics and photonics to the realm of phononics \cite{fang2016optical, MercierDeLepinay2020, Volz2016, Ma2019, Zhang2018, Ng2022}. In this context, cavity optomechanics provides a route towards the manipulation and control of phononic states via radiation-pressure forces \cite{Aspelmeyer2014, sanavio2020nonreciprocal, Barzanjeh2022}. Increasingly complex functionalities have been demonstrated in multimode optomechanical systems, where multiple optical and mechanical degrees of freedom interact \cite{xu2016topological, ruesink2018optical,xu2019nonreciprocal,Pelka2021}. 
Previously, several singular nonlinear phenomena have arisen from \textit{self-pulsing}, involving collective dynamics between free carriers and temperature in an optical cavity \cite{navarro2015self}. Such dynamics can produce rich dynamics, such as chaos \cite{navarro2017nonlinear},  injection-locking \cite{GArregui2021}, or frequency combs \cite{Allain2021}, but remain limited in frequency due to the relatively low thermalization rates encountered in these systems, which are typically on the order of a few tens of MHz. Here, we report an unexplored optomechanical dynamical regime that stems from the simultaneous self-sustained optomechanical oscillation of two mechanical modes that are far apart in frequency, where each mode independently interacts with a single optical mode. This mechanical-optical-mechanical (MOM) configuration is experimentally realized in a 2D membrane phononic crystal cavit y-waveguide sustaining both a 7 GHz phononic waveguide mode and a 265 MHz in-plane mechanical breathing mode of the full structure. Although thermal nonlinearities also exist in this optomechanical waveguide structure, the dynamics of the system is dictated purely by optomechanical interactions, which we confirm theoretically using a multimode optomechanical model. Consequently, the dynamics is set by mechanical properties of the structure that can be engineered over much shorter timescales than that of thermal effects. The multimodal interaction observed in this system leads to the formation and intermodulation of optical frequency combs in the GHz domain, with implications for atomic force and mass sensing \cite{Allain2021,Allain2020,Sansa2020,Yu2016}.

\begin{figure*}
\centering
    \includegraphics{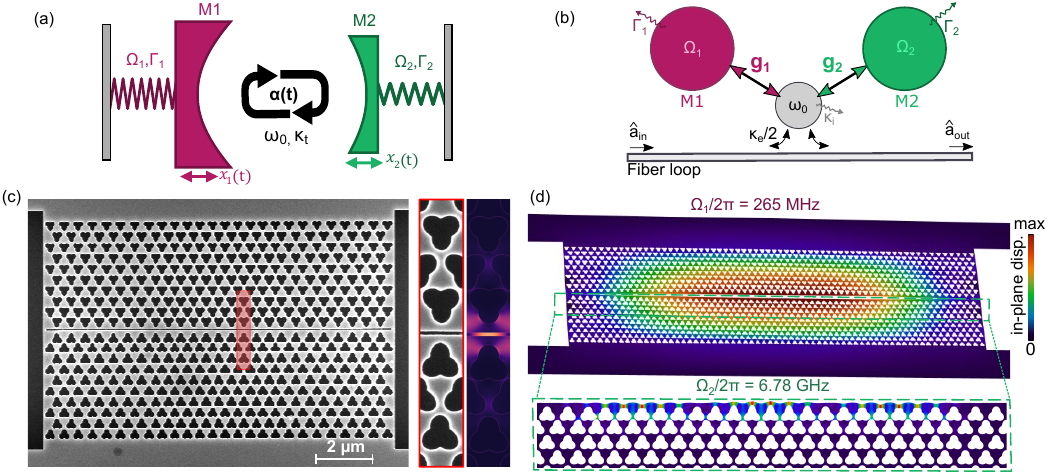}
    \caption{\textbf{Mechanical-optical-mechanical modal configuration.} (a) Generalized and (b) input-output schematic of a mechanical-optical-mechanical multimodal configuration, where two mechanical modes (\textit{M1} in red and \textit{M2} in green) couple to a waveguide-coupled single optical mode via their respective optomechanical coupling rates. The optical mode is not explicitly represented in (a), and is highlighted in grey in (b). (c) A scanning electron microscope image of the optomechanical experimental realization using suspended silicon nanomembranes. A zoom-in shows the slotted optical waveguide and a FDTD simulation of the electromagnetic energy confined within it. d) Displacement field distribution of the low-frequency in-plane mechanical mode \textit{M1} (top) and of the 3rd harmonic standing-wave acoustic mode traveling at the interface with the slot (bottom).}
    \label{Fig1}
\end{figure*}

\section{Model}

The archetypical representation of an optomechanical cavity uses an optical Fabry-Pérot cavity formed by two parallel mirrors, one of which has a mechanical degree of freedom, such as via attachment to a spring. \Cref{Fig1}a presents an adaptation of this representation to the MOM configuration, in this case using two moving mirrors. Each mirror (or mechanical mode) \textit{Mi} has its own natural mechanical frequency $\Omega_i$ and damping rate $\Gamma_i$, and couples to the optical field with a single-photon optomechanical coupling $g_i$. Their displacement is given by $x_i(t)$. An alternative input-output representation to more clearly observe inter-mode coupling is shown in \Cref{Fig1}b, where an optical waveguide (represented here by a fiber loop that is used in the experiment later on in this work) is side-coupled to an optical mode which interacts with \textit{M1} and \textit{M2}. The optical mode has a total decay rate  $\kappa_t=\kappa_i+\kappa_e$, where $\kappa_e$ is the decay rate associated with the input channel and $\kappa_i$ is the decay rate due to intrinsic losses. 
The linearized classical equations of motion for the cavity amplitude $\ac$ and for the mechanical displacements $x_1$ and $x_2$ can be expressed as:

\begin{equation}
\label{eq1}
    \begin{aligned}
        \dot{\ac} &= \Big(j\big(\Delta + g_1x_1+ g_2x_2\big) - \frac{\kappa_t}{2}\Big)\ac + \sqrt{\frac{\kappa_e}{2}}\ain \\
        \ddot{x_1} &= -\Gamma_1 \dot{x_1} -\Omega_1^2 x_1 + 2\Omega_1g_1|\ac|^2 \\ 
        \ddot{x_2} &= -\Gamma_2 \dot{x_2} -\Omega_2^2 x_2 + 2\Omega_2g_2|\ac|^2
        \end{aligned}
\end{equation}
where $\Delta=\omega_\ell-\omega_0$ is the detuning between the laser frequency $\omega_\ell=2\pi c/\lambda$ and the cavity resonance frequency $\omega_0$, and $\ain$ is the input laser field amplitude, such that $\pin=\hbar\omega_\ell|\ain|^2$ is the laser input power. Note that the displacements are normalized by each of the mechanical mode's zero-point fluctuation such that the mechanical masses can be eliminated. In a standard optomechanical cavity, the mechanical susceptibility, that describes the mechanical spectral response to an external perturbation, is altered by the laser field such that a \textit{direct} modification of both the mechanical frequency and damping occurs \cite{MarquardtPRL2007}. When considering a second mechanical mode, a correction term to each mechanical susceptibility can be derived (see \Cref{ap.calc}). This \textit{indirect} modification leads to gain competition between the two mechanical modes by acting against the direct optomechanical effect and therefore increases the threshold powers at which each mode can enter a self-sustained oscillation regime \cite{Kemitarak2014}. When $\Omega_1\sim\Omega_2$, simultaneous phonon lasing of both modes generally cannot occur, although it can be achieved using external modulation of the laser \cite{Mercade2021}.
This case of nearly degenerate mechanical modes implies that they couple to one another through light, leading to hybridization of the modes as observed in a SiN membrane placed inside an optical fiber cavity \cite{Shkarin2014}. The study of exceptional points \cite{miri2019exceptional}, which analyzes how gain and loss within a system can be balanced, has also been made possible using such a MOM configuration \cite{xu2016topological}, although the phonon lasing regime was not observed. When $\Omega_1\ll\Omega_2$, the coupling between \textit{M1} and \textit{M2} becomes non-resonant which allows for simultaneous lasing of the two mechanical modes at a relatively small input power and leads to the formation and intermodulation of combs that will subsequently be described.



\section{Experimental realization}

\begin{figure*}[!ht]
    \centering
    \includegraphics{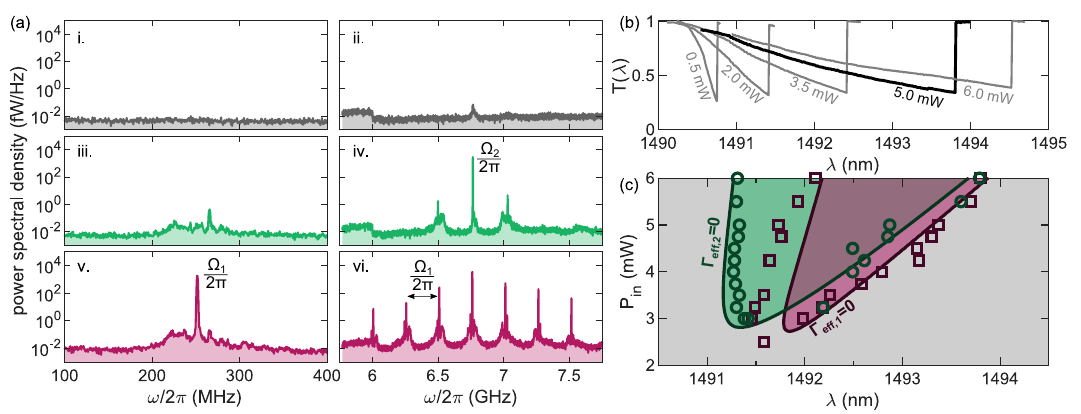}
    \caption{\textbf{Observed physical regimes.} (a) Mechanical RF spectra for the MHz range (i., iii. and v.) and GHz range (ii., iv. and vi.). Three characteristic dynamical regimes are observed: thermal (grey), \textit{M2} lasing (green), and comb intermodulation (red), which results from simultaneous \textit{M1} and \textit{M2} lasing. (b) Transmission response measured by driving a localized optical mode with a tunable laser from the blue-detuned side, using $\pin=4.5$ mW. (c) Theoretical map highlighting the lasing thresholds for \textit{M1} and  \textit{M2} as a function of the laser power and wavelength. The thresholds are shown with solid lines. The grey, green, and red areas correspond to both modes in the thermal regime, \textit{M2} lasing ($\Gamma_\mathrm{eff,2}<0$), and \textit{M1} lasing  ($\Gamma_\mathrm{eff,1}<0$), respectively. The experimentally determined thresholds of \textit{M1} (green circles) and \textit{M2} (red squares) are overlaid.}
    \label{Fig2}
\end{figure*}

While the underlying physics of such a multimode system are general, one example design in which the MOM configuration can be experimentally reproduced is in a 220 nm thick suspended silicon membrane fabricated in silicon-on-insulator with etched shamrock-shaped holes that form a two-dimensional phononic crystal (PnC). Two phononic crystals with mirror symmetry are brought together and separated by a thin air-slot within which light can be confined. \Cref{Fig1}c shows a scanning electron microscope (SEM) image of the PnC/air-slot/PnC platform with a zoom-in of the red shaded rectangle emphasizing the air-slot provided on the right. A finite element method simulation of the optical field energy density confined in the air-slot is also shown. Inherent fabrication imperfections such as etched sidewall roughness \cite{Albrechtsen2022} lead to disorder-induced spatially localized optical modes. From hereon, we are implicitly referring to these disorder-induced optical modes when mentioning optical modes in our structure. The localized modes provide efficient optomechanical transduction (i.e., the ability to couple light to mechanical motion) due to their high quality factors and small effective mode volumes, without the need to carefully design a confinement potential \cite{Arregui2021PRL}. The exact geometric parameters of the crystal can be found in \Cref{geom} and further details regarding fabrication and design of the optomechanical structure are found in Madiot, et al. \cite{Madiot2022PRL}. We define two mechanical modes observed in this structure as \textit{M1} and \textit{M2}, referring to the mechanical modes in the MHz and in the GHz regimes, respectively. 
In the structure explored here, \textit{M1} corresponds to the fundamental mechanical breathing mode in the MHz, or vibrations along the entire length of the plates, which strongly couples to the optical field in the slot, making it ideal for cavity optomechanical systems \cite{Li:10,Luan2014,safavi2010optomechanics,Arregui2021PRL}. \textit{M2} corresponds to a stationary wave resulting from the confinement of a mechanical guided mode in the acoustic Fabry-Pérot cavity formed on one side of the slot and between the two acoustic mirrors at the edges \cite{Madiot2022PRL,Florez2022}. The in-plane component of the displacement field associated with \textit{M1} and \textit{M2} are shown in \Cref{Fig1}d.
Both modes couple to the same optical mode, although they are unlikely to couple directly to one another as the overlap of the displacement fields is weak due to the large difference in length scale and frequency. By using a tapered-fiber loop brought into contact with the optical waveguide, these telecom wavelength optical modes can be driven to transduce both of these MHz and GHz mechanical modes. 

\section{Results and Discussion}

A localized optical mode is resonantly driven by injecting light from a tunable laser with input power $\pin=5.0$ mW in the fiber loop. The position of the loop for coupling to the mode and the polarization of the input light are optimized by maximizing the thermo-optic broadening of the mode, observed when scanning the lasing wavelength upward. The output field is analyzed with a fast photoreceiver connected to an electrical spectrum analyzer. As the optical mode is driven at increasing coupling-fractions by varying $\Delta$, different physical regimes are observed, summarized in their respective power spectral density spectra in \Cref{Fig2}a for the MHz (left panels) and the GHz (right panels). By approaching the optical mode with a blue-detuned laser, the first regime is a thermal regime in which both modes are thermally excited (grey). As the coupling mode fraction into the driven optical mode increases, a second regime appears in which a single GHz mechanical lasing peak is observed for \textit{M2} with side-bands formed by a thermally excited \textit{M1} in the MHz (green). For an even greater coupling fraction, a third regime appears in which lasing of both modes occurs (red). The exponentially growing oscillations that result from an optomechanical instability saturate due to nonlinear effects, which manifest as the emergence of various harmonics. The resulting spectral feature is an optical frequency comb \cite{Miri2018,Mercade2020}. The lasing of \textit{M1} in the MHz modulates the principal lasing peak of \textit{M2} in the GHz, giving rise to a frequency comb centered at $\Omega_{1}/2\pi=6.764$ GHz with spacing equal to $\Omega_{2}/2\pi=265$ MHz. 
The transmission optical spectrum associated with these measurements is shown in \Cref{Fig2}b (black line), accompanied by other scans at different input powers (grey lines). We observe additional broadening of the thermo-optically bistable resonance as $\pin$ increases.

By scanning the laser wavelength $\lambda$ over the resonance, we collect the lasing thresholds, or wavelengths at which each mode's amplitude begins an exponential increase or decrease, for both mechanical modes. We determine the mechanical lasing thresholds of \textit{M1} and \textit{M2} at different powers and report them in \Cref{Fig2}c with red squares and green circles, respectively. 
The optomechanical interaction alters the mechanical susceptibilities as computed via \Cref{eq1} such that the effective mechanical damping rates $\Gamma_\mathrm{eff,i}$ can be extracted to determine the phonon lasing condition of each mode. 
To allow for direct comparison to experiment, the theoretical reduced laser detuning $\Delta$ incorporates a resonant-wavelength $\lambda_0$ that is corrected for any thermo-optic shifts of the resonant cavity frequency using $\lambda_0^\prime = \lambda_0+\eta|\ac|^2$ where $\eta/\hbar\omega_\ell\approx0.35$  nm/fJ is a static nonlinearity estimated from the experimental data (see \Cref{klin}). The cavity photon number $|\ac|^2$ as a function of $\lambda$ and $\pin$ is numerically resolved by accounting for the static thermo-optic nonlinearity and the solutions are entered into the effective susceptibility to determine the effective damping rates. The calibrated experimental parameters are: $\lambda_0=1490.6$ nm, $\kappa_i/2\pi=5.0$ GHz, $\kappa_e/2\pi=4.5$ GHz, $\Gamma_1/2\pi = 1.5$ MHz, $\Gamma_2/2\pi = 3.4$ MHz, and $g_2/2\pi=260$ kHz.
\Cref{Fig2}c indicates when $\Gamma_\mathrm{eff,1}<0$ (green shaded region) and when $\Gamma_\mathrm{eff,2}<0$ (red shaded region). Each of these regions is delineated by the lasing threshold of each mode, $\Gamma_\mathrm{eff,1}=0$ and $\Gamma_\mathrm{eff,2}=0$. The experimental results are numerically fitted with the theory using $g_1$ as a fitting parameter. The best fit occurs for $g_1=437$ kHz and quantitatively captures the observed experimental values. This value is in good agreement with the optomechanical couplings found for the in-plane mechanical modes of similar slotted-waveguide optomechanical platforms \cite{Li:10,Grutter2015,Arregui2021PRL}.
The red region is mostly encompassed within the green one, representing a region in which simultaneous mode lasing occurs.

\begin{figure}[!ht]
    \centering
    \includegraphics{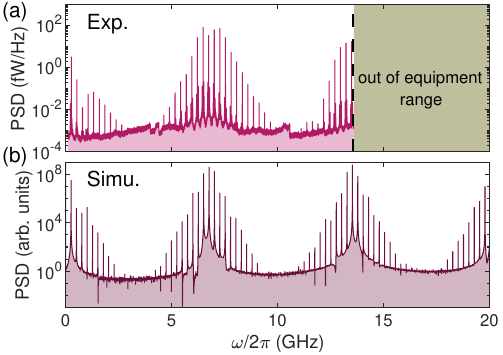}
    \caption{\textbf{Radiofrequency spectrum over full spectral range.} Power spectral density of the output optical field in the simultaneous lasing regime measured over the 0-13.5 GHz frequency range. Up to 10 pairs of sidebands on each side of $\Omega_2$ are observed, while the second harmonic of this pattern (centered at $2\Omega_2$) is partially visible near the signal analyzer limit.}
    \label{Fig3}
\end{figure}

\Cref{Fig3}a shows the power spectral density of the output optical field measured in the multimode lasing regime with a resolution bandwidth set to 50 kHz over the full span of the electrical spectrum analyzer from 0 to 13.5 GHz. \textit{M1} lases at 265 MHz, leading to the emergence of a first frequency comb with about 10 harmonics up to 2.650 GHz. \textit{M2} also lases, but the second harmonic already lies above the equipment limit. The harmonics of \textit{M1} are imprinted in the frame rotating at $\Omega_2$, such that about 10 pairs of sidebands are visible around $\Omega_2$. This frequency pattern is duplicated in all the harmonics of \textit{M2}, such that the beginning of its second occurrence (centered at $2\Omega_2$) can be spotted above 12 GHz. 
The experimental spectrum is plotted in the range 0-20 GHz for direct comparison with \Cref{Fig3}b which shows a numerical simulation. It shows the same type of spectral structure with slight variations of the harmonic amplitude distribution. 

Other regimes involving another MHz mechanical mode as well as self-pulsing (SP) dynamics could also be observed within this structure for certain experimental input parameters but are ignored by our model. In particular, a SP regime with thermal oscillation at $\Omega_\mathrm{SP}/2\pi=13.5$ MHz is found in addition to simultaneous optomechanical lasing of \textit{M1} and \textit{M2}, leading another level of intermodulation (\Cref{SPregime}). 

\begin{figure*}[!ht]
    \centering
    \includegraphics{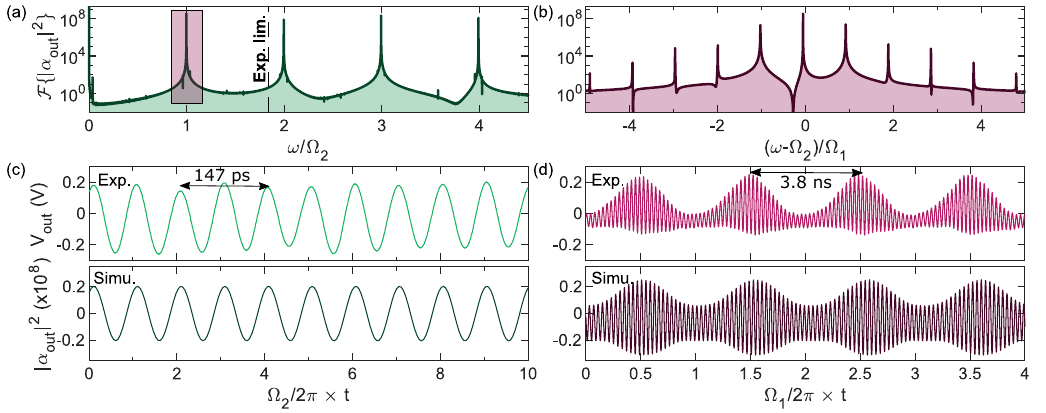}
    \caption{\textbf{Physical regimes in the time domain.} Simulated Fourier spectrum of $|\aout(t)|^2$ for (a) the \textit{M2} lasing regime and (b) the comb intermodulation arising from simultaneous \textit{M1} and \textit{M2} lasing. The experimental measurement cut-off due of the measurement set-up is indicated with a dashed line in (a). For comparison, the frequency range from (b) is highlighted in (a) by a red box. Experimental (top) and simulated (bottom) time traces are shown for the (c) \textit{M2} lasing regime and (d) multimode lasing regime.}
    \label{fig4}
\end{figure*}

The regimes where \textit{M2} lases independently (\Cref{Fig2}a iii-vi) and where both modes lase simultaneously (\Cref{Fig2}a v-vi) are reproduced numerically using \Cref{eq1}, and are shown in \Cref{fig4}a and b, respectively, using the output photon flux $|\aout(t)|^2$ where $\aout = \ain-\sqrt{\frac{\kappa_e}{2}}\ac$ (see \Cref{NumericsMethodology}). 
The experimental oscilloscope cut-off frequency is at 8 GHz. 
The associated experimental (top) and simulated (bottom) time traces for each case are shown in \Cref{fig4}c and d. 
Our experimental apparatus is limited to 13.5 GHz (dashed line in \Cref{fig4}a) which lies slightly below the second harmonic. Both of the associated time traces for this scenario (\Cref{fig4}c) show slightly anharmonic oscillation at $\Omega_2$. A low-pass filter with cut-off frequency at 8 GHz is applied to the simulation time traces (bottom) to reproduce the oscilloscope bandwidth. 
For the case where simultaneous lasing occurs (\Cref{fig4}b), the frequency is centered at $\Omega_2$ and normalized with $\Omega_1$, indicating the presence of a comb centered at $\Omega_2$ with spacing given by $\Omega_1$. This spectrum shows three pairs of sidebands, which are also observed experimentally in \Cref{Fig2}a (vi). These correspond to the first four harmonics of \textit{M1} that imprint in the frame rotating at $\Omega_2$ via the indirect nonlinear coupling of the mechanical modes. These harmonics emerge when \textit{M1} passes its lasing threshold, similar to the case of \textit{M2}. 
The experimental time-trace in \Cref{fig4}d (top) shows amplitude-modulation of the GHz tone by the MHz tone, in the output optical field. The simulated trace (bottom) exhibits good qualitative agreement with experiment.

\section{Conclusion}

We presented a mechanical-optical-mechanical multimodal optomechanical platform, connecting spectrally distant phonons and photons in the MHz, GHz, and hundreds of THz, and demonstrated optical control the mechanical dynamics. Our platform exhibits the intermodulation of two optical frequency comb in the GHz domain, resulting from self-sustained mechanical oscillations of a GHz PnC guided mode that interacts with a lasing MHz breathing mode via the optical field, which occurs independent of thermal effects, as supported by our theoretical model. 

The modulation frequency ($\Omega_1$) can also be actively tuned even further via an external stimulus, such as an atomic force microscope tip  or other static actuator which acts as a local perturbation with the dampening leading to an order of magnitude energy loss \cite{Rieger2014}, or with additional electrostatic actuation  as envisioned with electro-optomechanical systems \cite{midolo2018nano,Pitanti:15}.
In our case, this dampening would significantly shift \textit{M1} and the modulation frequency by consequence. With regards to the spectral line shape, Miri et al. \cite{Miri2018} demonstrated that this distribution can be readily controlled via the optical losses of the cavity, which can also be engineered and experimentally tuned. Such  modulation scheme further enhances understanding and versatility of high-frequency input-output phononic devices \cite{yu2016surface, PhysRevLett.128.015501}. 

More generally, such a process enriches the level of control in multimode optomechanical platforms and could be used for applications such as realizing sub-threshold multimode lasing of two GHz modes without the need for an external intermodal modulation of the laser \cite{Mercade2021}. Within this context, the extent to which the modulation frequency as well as the spectral distribution of the frequency comb lines can be readily controlled has yet to be studied, though such prospects could lead towards the realization of optomechanical logic gates \cite{Pelka2021}. 




\begin{acknowledgements}
We acknowledge the support from the project LEIT funded by the European Research Council, H2020 Grant Agreement No. 885689. ICN2 is supported by the Severo Ochoa program from the Spanish Research Agency (AEI, grant no. SEV-2017-0706) and by the CERCA Programme / Generalitat de Catalunya. R.C.N. acknowledges funding from the EU-H2020 research and innovation programme under the Marie Sklodowska Curie Individual Fellowship (Grant No. 897148). O.F. is supported by a BIST PhD fellowship under the Marie Sklodowska Curie grant agreement (No. 754558). M.A. and S.S. gratefully acknowledge funding from the Villum Foundation Young Investigator Program (Grant No.\ 13170), the Danish National Research Foundation (Grant No.\ DNRF147 - NanoPhoton), Innovation Fund Denmark (Grant No.\ 0175-00022 - NEXUS), and Independent Research Fund Denmark (Grant No.\ 0135-00315 - VAFL).
\end{acknowledgements}

\appendix

\section{Derivation of the mechanical susceptibilities}
\label{ap.calc}

The presence of a second mechanical mode leads to a deviation in the effective susceptibility of each mechanical oscillator. The natural mechanical susceptibility of oscillator \text{i} is $\chi_i^{-1}(\omega) = \Big[ (\Omega_i^2 - \omega^2) - j\omega\Gamma_i\Big]$. A modification to this lorentzian response arises due to the optomechanical interaction, as calculated in \cite{Aspelmeyer2014}, with $\Sigma_i(\omega) = 2\Omega_ig_i^2\beta(\omega)$ where $\beta(\omega) = \frac{\ncav}{(\Delta+\omega)+j\kappa_t/2}+\frac{\ncav}{(\Delta-\omega)-j\kappa_t/2}$ with $\ncav=|\ac|^2$ the cavity photon number.
Applying the same analytical treatment to the MOM configuration, the linearization of \Cref{eq1} leads to:
\begin{equation}
\label{eq2}
    \begin{aligned}
        \big[\chi_1^{-1}(\omega) + \Sigma_1(\omega)\big]x_1 = 2\Omega_1g_1g_2\beta(\omega)x_2\\
        \big[\chi_2^{-1}(\omega) + \Sigma_2(\omega)\big]x_2 = 2\Omega_2g_1g_2\beta(\omega)x_1
        \end{aligned}
\end{equation}
This expression highlights that the coupling between the mechanical modes scales with $g_1g_2$ which is expected since they do not interact directly but through the optical field.
Note that it is almost equivalent to the Hamiltonian formulation presented in \cite{xu2016topological}, where the authors additionally assume nearly-degenerate modes, which is not the case in the present work.
Solving \Cref{eq2} provides an expression for the indirect coupling term between the mechanical modes:


\begin{equation}
K_i(\omega) = -\frac{\Sigma_i(\omega)\Sigma_j(\omega)}{\chi_j^{-1}(\omega) + \Sigma_j(\omega)}
\end{equation}

such that the effective mechanical susceptibility of the oscillator \textit{i} becomes:

\begin{equation}
\label{totSusc}
    \chi_\mathrm{eff,i}^{-1}(\omega) = \chi_i^{-1}(\omega) +\Sigma_i(\omega)+ K_i(\omega)
\end{equation}

From \Cref{totSusc}, we evaluate the effective mechanical damping rates $\Gamma_\mathrm{eff,i}=-\mathrm{Im}[\chi_\mathrm{eff,i}]/\Omega_i$ in the parameter space $\{\lambda,\pin\}$ to build the stability diagram presented in \Cref{Fig2}c.
We note that the indirect coupling term becomes increasingly small as the frequencies $\Omega_1$ and $\Omega_2$ are moved apart. The case of nearly-degenerate modes $\Omega_1\sim\Omega_2$ is beyond the present scope.
 
\section{Geometrical parameters}
\label{geom}

\begin{figure}[!ht]
    \centering
    \includegraphics[scale = 0.28,clip]{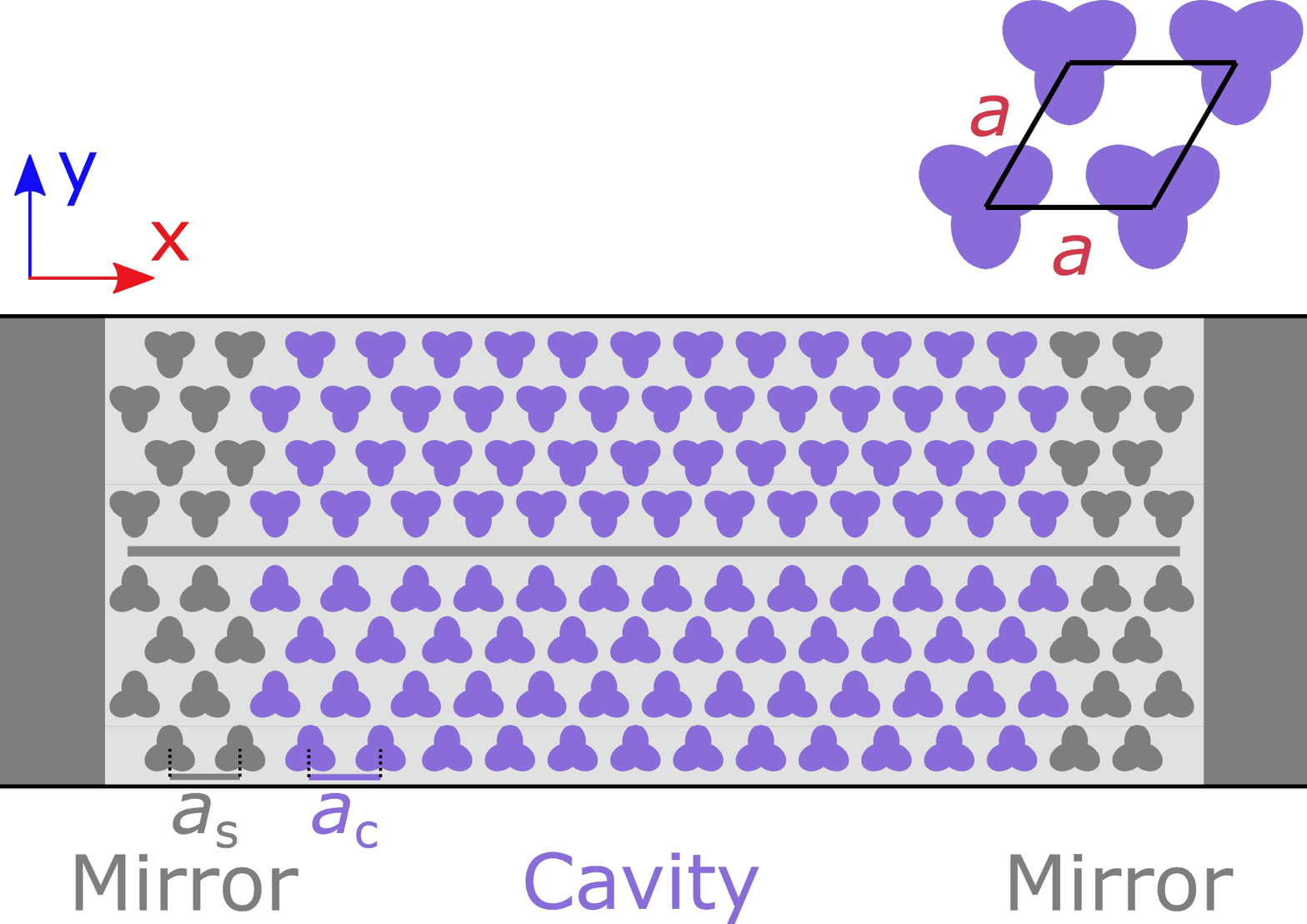}
    \caption{\textbf{Schematic of waveguide structure.} A triangular lattice of \textit{shamrock} shaped holes form the crystal (top). Dark grey indicates etched air holes while light grey indicates remaining silicon material. The phononic crystal is formed by a center cavity of etched shamrock holes (highlighted in light purple) and surrounding acoustic mirror regions. Two identical plates are placed with opposing y-symmetry across from one another seprated by a rectangular slot-waveguide within which light can localize.}
    \label{Fig_geom}
\end{figure}

The shamrock motifs are formed by three overlapping elliptical holes, with each ellipse defined by short and long axes given by $n_a = 0.23a_c$ and $n_b = 0.30a_c$, respectively, where $a_c$ = 500 nm is the center PnC lattice period (\Cref{Fig_geom}). This schematic is not to scale. Light grey indicates the remaining, unetched, suspended silicon material. The center PnC cavity region (light purple, $a_c$ = 500 nm) is surrounded on both sides by separate mirror PnC regions with larger period (dark grey, $a_s$ = 560 nm). The length of the center waveguide region can be varied with the number of periods. In the fabricated sample, the mirror region has 10 periods in the x-direction, while both the mirror and cavity regions have 10 periods in the y-direction on each side of the slot to ensure confinement of the acoustic wave. The row of shamrocks closest to the center air-slot surface are shifted away from the edge by $d_y\sqrt{3}a_c/2$, where $d_y$ = 0.1155. Subsequent rows along the y direction are kept at their nominal positions in the crystal lattice. A second phononic crystal with opposite y-symmetry is brought near the original and the two are separated by a narrow air slot whose width is also set by $d_y$. 

\section{Thermo-optic effect and free-carrier absorption}
\label{klin}

\begin{figure}[!ht]
    \centering
    \includegraphics[scale = 1]{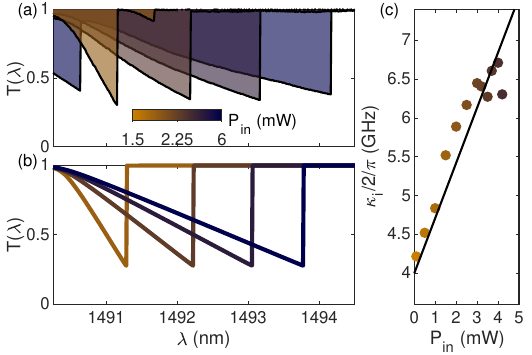}
    \caption{\textbf{Thermo-optic effect and free-carrier absorption}. (a) Normalized transmission response for increasing input power (color bar) leading to a thermo-optic shift of the optical resonance. (b) Calculated solutions of the transmission using a static thermo-optic nonlinearity $\eta/\hbar\omega_\ell\approx0.35$  nm/fJ, plotted for increasing input power. (c) Input power dependence of the intrinsic loss rate calculated with the minimum transmission $T_0$ extracted from (a)using $\kappa_i = \kappa_e \sqrt{T_0}/(1-\sqrt{T_0})$. Data are fitted with a linear function (solid line).}
    \label{Fig_klin}
\end{figure}

Increasing the input laser power $\pin$, the cavity resonance wavelength experiences a shift due to the thermo-optic effect. \Cref{Fig_klin}a shows the normalized transmission response measured at different input powers. The wavelength is swept upward from the blue-detuned side which reveals the hysteretic behaviour of the optical resonance. At low power, a second resonance dip slightly couples with the fiber loop although it is red-shifted away from the main resonance for higher power.

The intracavity photon number $|\ac|^2$ is calculated using a static thermo-optic nonlinearity $\eta/\hbar\omega_\ell\approx0.35$  nm/fJ with a numerical solver. The associated transmission function $T(\lambda)$ is plotted in \Cref{Fig_klin}b for increasing power. 

The minimum transmission value obtained by sweeping the wavelength only depends on the internal and external decay rates, $\kappa_i$ and $\kappa_e$ respectively, and is given by $T_0=\kappa_i^2/\kappa_t^2$. The internal decay rate can then be experimentally deduced using $\kappa_i = \kappa_e \sqrt{T_0}/(1-\sqrt{T_0})$. When increasing the intracavity energy, two-photon absorption by the free-carriers leads to an increase in $\kappa_i$, as shown in \Cref{Fig_klin}c with the data extracted from \Cref{Fig_klin}a (dots). This trend is fitted with a linear function (black line). The internal loss rate increases by a factor $\sim1.6$ between 0 and 6 mW.

\section{Self-pulsing and other multimode lasing regimes}
\label{SPregime}

\begin{figure}
    \centering
    \includegraphics{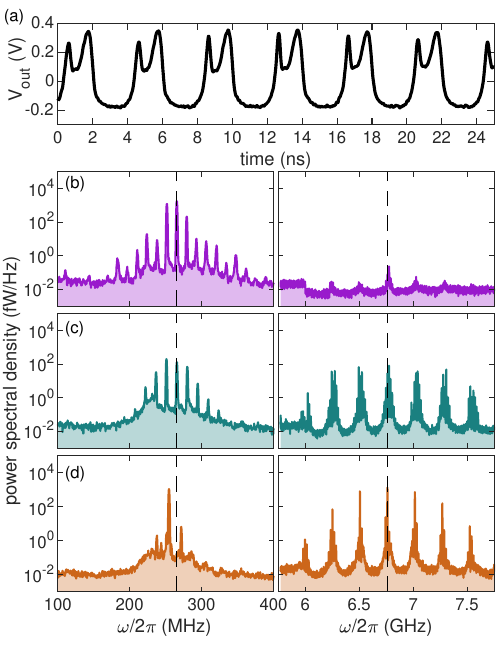}
    \caption{\textbf{Additional dynamical regimes.} (a) Experimental time trace recorded in the SP regime in the absence of optomechanical lasing. Power spectral density spectra measured for different dynamical regimes with (b) SP oscillations and \textit{M1} optomechanical lasing leading to a comb centered at $\Omega_1/2\pi$ and with spacing $\Omega_\mathrm{SP}/2\pi$, (c) SP oscillations and simultaneous \textit{M1} and \textit{M2} optomechanical lasing, leading to the nesting of a frequency comb within another comb, and (d) simultaneous optomechanical lasing of \textit{M2} and another MHz mechanical mode with frequency slightly lower than $\Omega_1/2\pi$. For reference, the dashed lines indicate $\Omega_1/2\pi$ and $\Omega_2/2\pi$ in the left and right panels, respectively.}
    \label{Fig_SP}
\end{figure}

Several other dynamical regimes have been identified in the power spectral density of the output optical field. These regimes are beyond the scope of the present manuscript as they involve self-pulsing (SP) dynamics or an additional mechanical mode. SP is a self-sustained oscillation regime resulting from the competition between thermalization processes and free-carrier relaxation in the optical cavity. The SP oscillation is nonlinear, i.e. anharmonic, and leads to a series of spectral peaks in the frequency domain. \Cref{Fig_SP}a shows a time-trace measured in a SP regime, in the absence of any optomechanical oscillations such that the fast carrier response followed by a slower thermal oscillation can be clearly identified. The trace displays nonlinear oscillation with fundamental frequency $\Omega_\mathrm{SP}=13.5$ MHz.

The combination of such SP oscillation with optomechanical lasing of \textit{M1} leads to a frequency comb in the MHz domain, intermodulating the one centered at $\Omega_1$, as illustrated in \Cref{Fig_SP}b. \textit{M2} is not lasing for this case. This regime is found for $\pin=4.75$ mW, around $\lambda=1492.24$ nm.
In \Cref{Fig_SP}c, both \textit{M1} and \textit{M2} are now lasing, while the SP oscillations persist. This results in the nesting of three frequency combs in the GHz domain involving $\Omega_1$, $\Omega_2$, $\Omega_\mathrm{SP}$, and their harmonics. This regime is found for $\pin=4.75$ mW, around $\lambda=1493.14$ nm.
While other works have treated the SP dynamics \cite{navarro2015self,Allain2021}, the observation of a purely mechanical (i.e., without thermal effects) frequency comb appearing in the GHz regime based on a MOM configuration has not been previously reported. 
Finally, in \Cref{Fig_SP}d, we report another regime where two mechanical modes lase simultaneously. However, while the GHz mode is still represented by \textit{M2}, an additional MHz mode appears that is different from \textit{M1}, with frequency around 254 MHz. This regime is found for $\pin=4.75$ mW, around $\lambda=1492.79$ nm.
This work focuses on the combined optomechanical dynamics of \textit{M1} and \textit{M2}, although these additional regimes can overtake the multimode lasing dynamics. For these special cases, there exist certain points in the parameter space at which the exact lasing threshold in \Cref{Fig2}c is difficult to determine.

\section{Numerical simulations}
\label{NumericsMethodology}
The numerical simulations are performed with an adaptive step-size fourth-order Runge-Kutta simulation of \Cref{eq1}. The simulation is realized ignoring any thermal nonlinearity ($\eta=0$), using the calibrated parameters, and setting the input power to $35$ mW. Lower values also reproduce the observed dynamics, but the absence of the thermo-optic nonlinearity makes the spectral window in which it occurs very narrow  (few picometers), and the system tends to diverge from the optomechanical bistability excited state after a few 100s of nanoseconds.
The wavelength $\lambda$ is increased from off-resonant to resonant driving to converge to the high-energy state of the optomechanical bistability. The simulation in the single-mode lasing regime (\Cref{fig4}c) is obtained at $\lambda=1490.520$ nm with initial conditions $\{ \ac,\dot{x_1},x_1,\dot{x_2},x_2\}=\{2.681\times10^2 + 1.523\times10^2i, -9.784\times10^{10},3.480\times10^3, -3.784\times10^{15}, 3.899\times10^4\}$. The simulation in the multimode lasing regime (\Cref{fig4}d) is obtained at $\lambda=1490.564$ nm with initial conditions $\{ \ac,\dot{x_1},x_1,\dot{x_2},x_2\}=\{1.326\times10^3 - 4.280\times10^2i, -1.965\times10^{12}, -1.681\times10^3, -1.751\times10^{15}, -6.173\times10^4\}$.


\clearpage
\bibliography{MOM_biblio}

\end{document}